\documentclass[]{spie} 

\usepackage{amsmath,amsfonts,amssymb}
\usepackage{graphicx}
\usepackage[colorlinks=true, allcolors=blue]{hyperref}
\usepackage{subcaption}
\usepackage{siunitx}
\usepackage{hhline}
\usepackage{array}
\usepackage{placeins}

\newcommand{\RNum}[1]{\uppercase\expandafter{\romannumeral #1\relax}}

\title{Silicon drift detectors for the Spectroscopy Focusing Array of eXTP}

\author[a]{A. Altmann}
\author[a]{T. F. Bechteler}
\author[a]{R. Strecker}
\author[b]{P. Lechner}
\author[a]{R. Andritschke}
\author[a]{G. Hauser}
\author[c]{C. Fiorini}
\author[a]{K. Nandra}
\affil[a]{Max-Planck-Institut für extraterrestrische Physik, Gießenbachstr. 1, 85748 Garching, Germany}
\affil[b]{Halbleiterlabor der Max-Planck-Gesellschaft, Isarauenweg 1, 85748 Garching, Germany}
\affil[c]{Politecnico di Milano, Piazza Leonardo da Vinci 32, 20133 Milano, Italy}

\authorinfo{Alexander Altmann: E-mail: aaltmann@mpe.mpg.de, Telephone:  +49 (0) 89 30 000 3225}

\begin{document} 
\maketitle

\begin{abstract}
	We present a silicon drift detector (SDD) system for the spectroscopy focusing array (SFA) of the enhanced X-ray timing and polarimetry (eXTP) mission. The SFA focuses on fast timing (time resolution below $ \SI{10}{\micro\second} $) and good spectroscopy capabilities (energy resolution better than 180 eV @ 6 keV). The sensor, consisting of 19 hexagonally shaped pixels with a total sensitive area of 5.05 cm², is connected to three high time resolution spectroscopy (HTRS) ASICs, allowing a fast readout of the detector signals. The detector works in a Charge-Sensitive Amplifier configuration. We assembled a prototype detector module and present here its mechanical design, describe the used sensor, and report about its performance. 
\end{abstract}


\section{INTRODUCTION}
\label{sec:intro} 

The eXTP (enhanced X-ray Timing and Polarimetry) mission is a space mission led by the Institute of High Energy Physics (IHEP) of the Chinese Academy of Sciences (CAS) with contributions from European partners. Its goal is to explore the physics of ultra-dense matter and the behavior of matter under strong gravitational and magnetic forces \cite{Zhang2018}. The eXTP satellite features four main instruments: The wide field monitor (WFM), the large area detector (LAD), the polarimetry focusing array (PFA), and the spectroscopy focusing array (SFA).
The SFA consists of nine telescopes, each has a Wolter-\RNum{1} mirror focusing X-rays onto a detector module. 
We chose a Silicon Drift Detector (SDD) for the SFA detector.
The concept of a Silicon Drift Detector was introduced by E. Gatti and P. Rehak, describing the silicon drift chamber \cite{Gatti1984}. An advantage of the SDD is the small readout anode and therefore low readout capacitance, independent of the sensor size. This leads to a good signal-to-noise ratio and improved performance, even at high count rates. Therefore, SDDs are well suited for applications in X-ray astronomy.
In order to fully profit from the small capacitance, the first FET is often integrated directly on the sensor \cite{Lechner1996}. The sensors for eXTP SFA also have a FET integrated in the center of every pixel.
An example of another space mission that uses SDDs is the NICER mission. \cite{Gendreau2016,LaMarr2016} 
\newline
The eXTP satellite shall observe very bright X-ray sources with intensities of up to 20 Crab, which 
requires a high detector throughput. 
Referring to the eXTP SFA requirements, the detector dead time must be lower than 5 \% of the total observation time when a 1 Crab X-ray source is observed.
Considering the mirrors of SFA, 1 Crab is equivalent to a count rate of approximately 6.4 kcps at the center pixel.
The energy resolution at 6 keV is required to be better than 180 eV, and the energy range in which SFA shall operate is $ \SI{0.5}{\kilo\electronvolt} $ to $ \SI{10}{\kilo\electronvolt} $ \cite{Zhang2022}.
In this work, we present the mechanical design of the prototype detector module for SFA. Afterwards, the detector is described in more detail, and how the sensor works together with three High Time Resolution Spectroscopy (HTRS) ASICs.
Finally, measurements of the detector performance are presented, focusing on its energy resolution and leakage current.

\section{mechanical design of the detector module}
\label{sec:desi_mod}  
\begin{figure} [ht]
	\begin{center}
		\begin{tabular}{c} 
			\includegraphics[height=10cm]{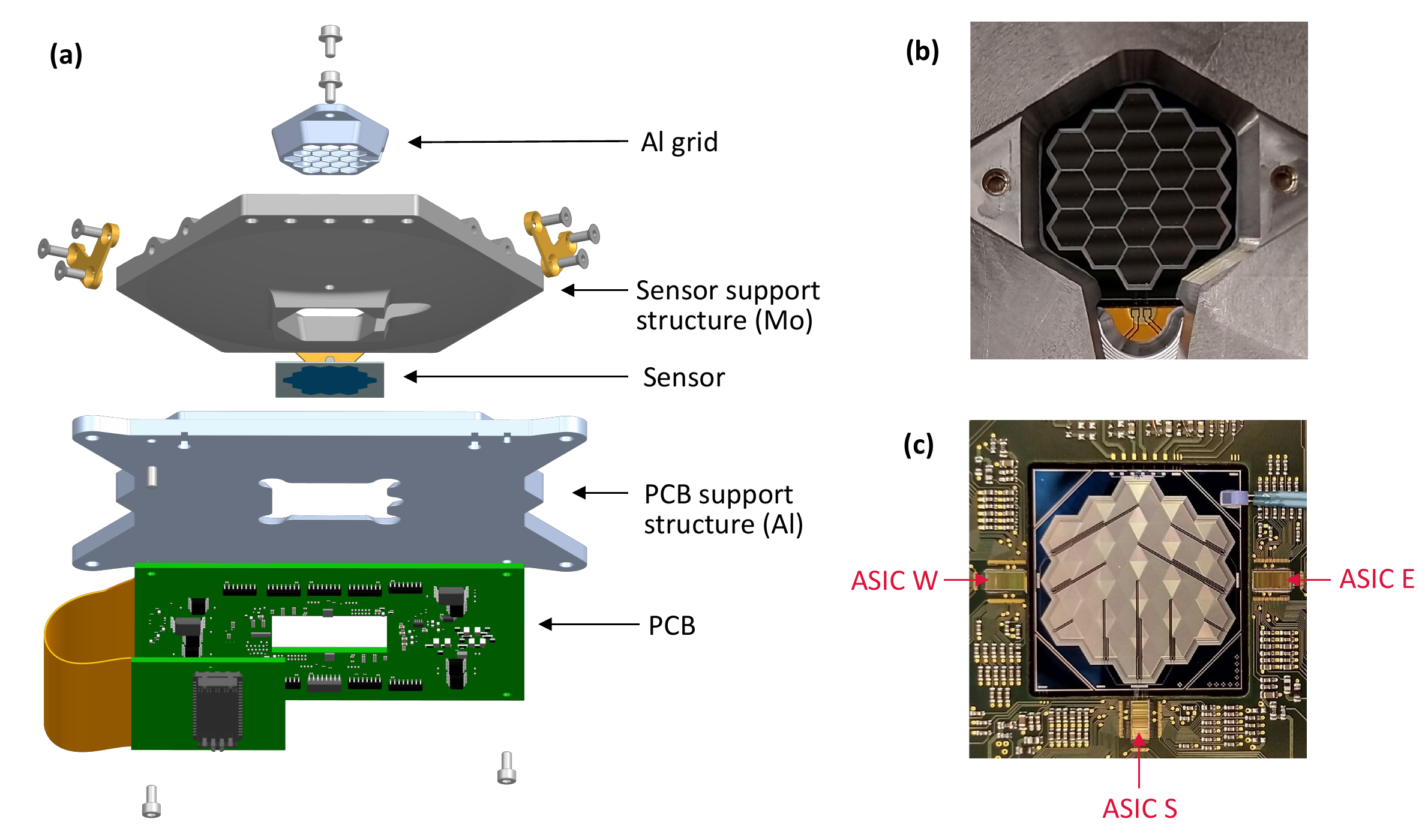}			
		\end{tabular}
	\end{center}
	\caption[example] 
	{ \label{fig:module_mechanics} 
		\textbf{(a)} Detector module consisting of two metal support structures that are connected with PEEK struts (golden colour) for thermal insulation. The silicon sensor is glued onto the upper support structure made of molybdenum, the lower support structure is made of aluminium and carries the PCB with the FEE. \textbf{(b)} Entrance window side of the 19-pixel silicon drift detector in the center of the detector module. \textbf{(c)} Front side of the sensor. The three HTRS ASICs used for the detector readout are indicated. Each pixel of the sensor is connected to one HTRS ASIC channel via aluminium bond wires.}
\end{figure}

The detector module's mechanical design is shown in Figure \ref{fig:module_mechanics}(a). It consists of two separate metal support structures. 
The upper one is made of molybdenum, the SDD sensor is glued to its center.  
Molybdenum is chosen as a material since the thermal expansion coefficient is close to that of silicon to avoid unnecessarily large strain on the sensor due to temperature changes.
The lower support structure made of aluminium carries the printed circuit board (PCB). This PCB contains the front-end-electronics (FEE), most importantly the three HTRS ASICs used to read out all of the 19 pixels. 
These HTRS ASICs were designed at the Politecnico di Milano in Italy \cite{Quaglia2014}. 
For thermal considerations, the two support structures are separated by very weakly thermally conducting Polyether ether ketone (PEEK)-struts. This allows the upper support structure, including the sensor, to be cooled down to temperatures as low as $ \SI{-30}{\degreeCelsius} $, while the lower support structure is connected to the satellite structure that is passively cooled and kept at a temperature around $ \SI{20}{\degreeCelsius} $. 
Since the heating power of the FEE with the HTRS ASICs as main contributors
is approximately ten times as large as the heating power of the sensor, less cooling power is then required to keep the sensor at the desired low temperature.
The metal surfaces have to be polished to reduce the thermal emittance of the aluminium part and also reduce the thermal absorptance of the molybdenum part while increasing its reflectance. 
We are currently working on reducing the overall area of the two support structures to further decrease the heat transfer due to radiation between the bottom and top metal plates.
This transfer is significant since the two plates have a big surface area parallel to each other. \newline
The total input of heating power on the molybdenum part due to thermal radiation was estimated to approximately $ \SI{1.1}{\watt} $. This contains thermal radiation from the aluminium part as well as the satellite surroundings. 
Additionally, thermal conduction via the bond wires between the sensor and ASICs, as well as via the PEEK struts between the aluminium and molybdenum support structure, is present with a power of $ \SI{0.3}{\watt} $.
The biasing of the SDD sensor causes a heating power of $ \SI{0.1}{\watt} $. Combining all contributions, the resulting heating power of $ \SI{1.5}{\watt} $ has to be dissipated to keep the sensor's temperature constant.
The sensor is cooled via the molybdenum structure, which will contain mechanical interfaces to connect thermal straps to the edges. The plan is to use thermoelectric coolers (TEC) that can be placed on the other side of the thermal straps to create the desired temperature gradient between the sensor and the radiator.
To test the detector performance in the laboratory, we use a stirling cooler connected to the molybdenum support structure to cool down the module. 
Preliminary tests were performed with a dummy thermal model consisting of a molybdenum plate and a steel plate.
A TEC with an electrical power of $ \SI{10}{\watt} $ is sufficient to cool the molybdenum plate to a temperature of $ \SI{-30}{\degreeCelsius} $, whereas the warm TEC side is at $ \SI{-8}{\degreeCelsius} $. 
With such a TEC, the power the radiator has to dissipate would then add up to approximately 11.5 W.
In the experiments, ohmic resistors were used to simulate the power dissipation of the HTRS ASICs and the SDD sensor.
The temperature difference over the sensor area was roughly $ \SI{1}{\degreeCelsius} $. 
It is expected that both the required electrical power of the TEC and the temperature gradient over the sensor will decrease once the module redesign is finished.
In order to monitor the temperatures of the SDD and the PCB, several temperature sensors are mounted on the module. A resistance thermometer of the type PT1000 can be seen in Figure \ref{fig:module_mechanics}(c) in the top right corner of the sensor chip. 
There is also a temperature diode integrated on the sensor itself that can be used to determine its temperature. \newline
An aluminium grid is provided to be mounted onto the molybdenum structure from the top (photon entrance side). 
It covers the sensor pixel edges and shadows the areas where photons generate split events. A split event is described as the sharing of signal charges between neighbouring pixels.

\section{Detector Design}
\label{sec:desi_det}

\begin{figure} [h!]
	\begin{center}
		\begin{tabular}{c} 
			\includegraphics[height=6.67cm]{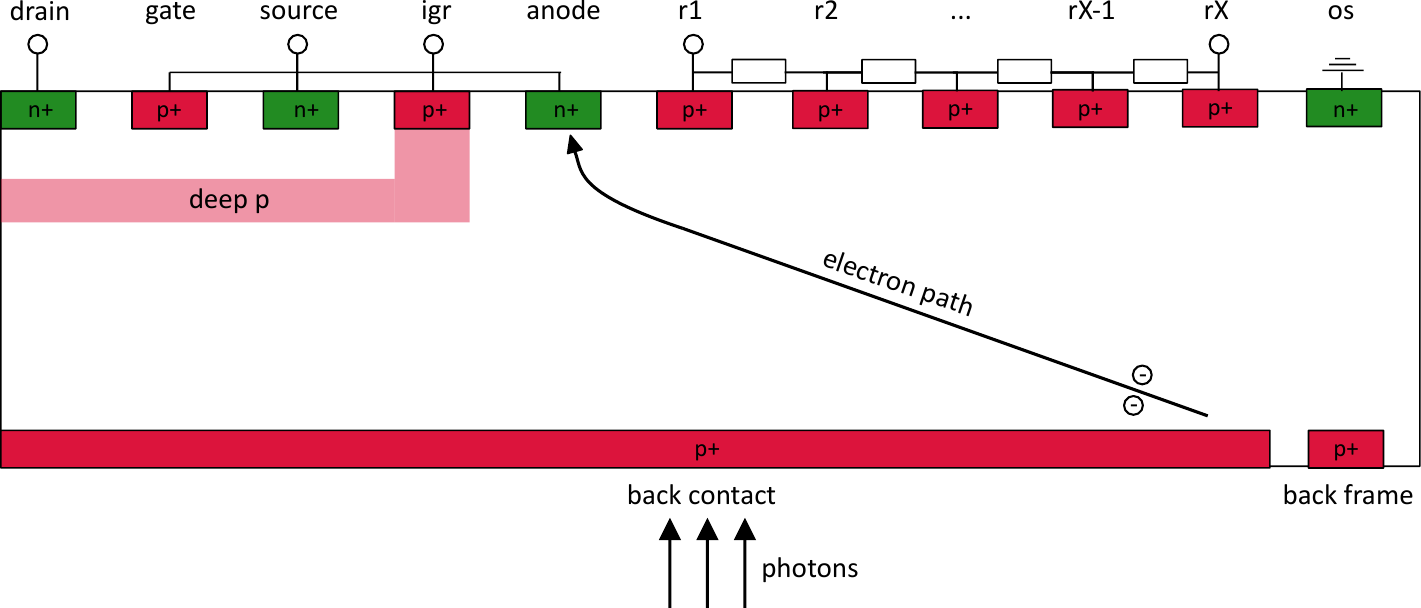}
		\end{tabular}
	\end{center}
	\caption[example] 
	{ \label{fig:sensor_scheme} 
		Schematic cross-section of a sensor pixel from the center (drain contact) to the pixel edge (outer substrate contact os). The integrated FET in the pixel center is shown, as well as the drift rings (r1-rX).  The total number of drift rings is 40. Proper biasing of the drift rings via the integrated voltage divider guarantees that the generated electrons drift to the anode on the drawn path. The wafer has a thickness of $ \SI{450}{\micro\meter} $.}
\end{figure}

\begin{figure} [ht]
	\begin{center}
		\begin{tabular}{c} 
			\includegraphics[width=0.95\linewidth]{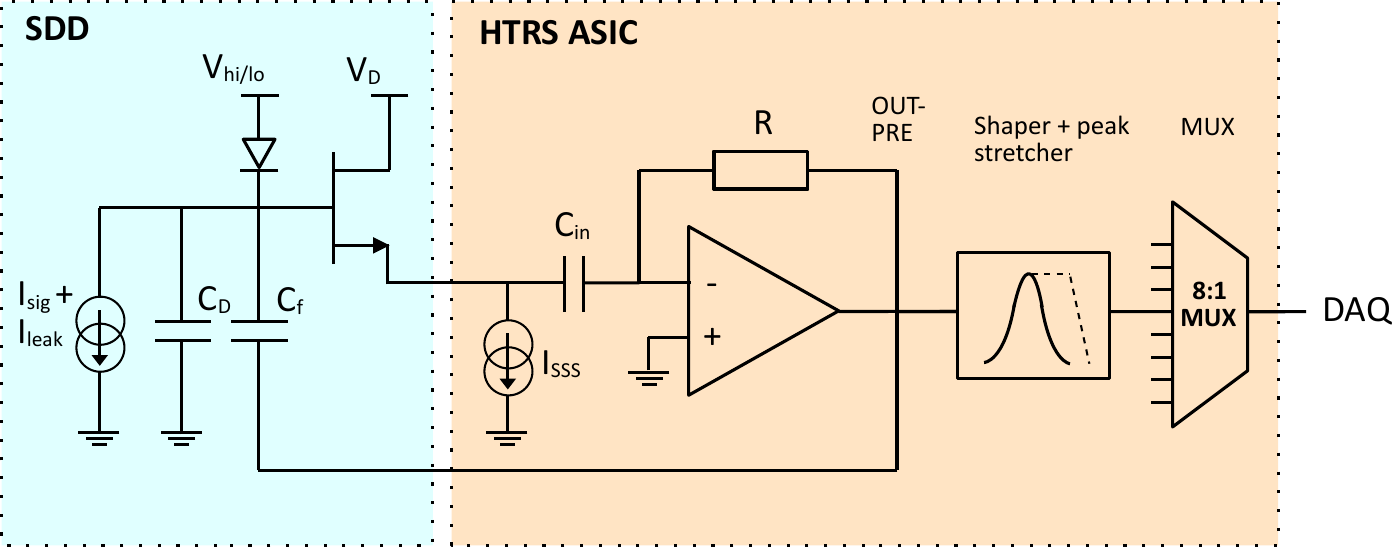}
		\end{tabular}
	\end{center}
	\caption[example] 
	{ \label{fig:readout} 
		Detector readout circuit showing one SDD sensor pixel and a corresponding channel of the readout HTRS ASIC. The output of the first preamplifier located in an HTRS ASIC channel is connected to a sensor pixel via a feedback capacitance, forming a charge-sensitive amplifier. }
\end{figure}

\begin{figure} [ht]
	\begin{center}
		\begin{tabular}{c}
			\includegraphics[width=0.95\linewidth]{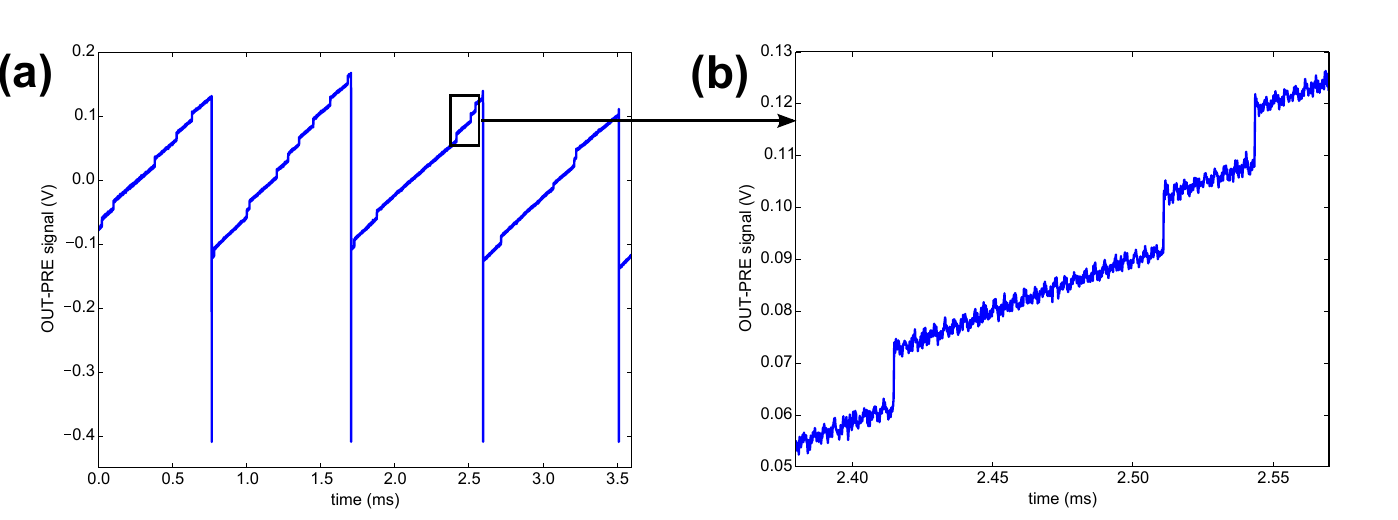}
		\end{tabular}
	\end{center}
	\caption[example] 
	{\label{fig:outsig}\textbf{(a)} Monitoring voltage signal of the CSA in channel S8 at a sensor temperature of $ \SI{0}{\degreeCelsius} $ (see Figure \ref{fig:enres}(c) for the location of the pixel connected to S8 on the sensor). The constant positive slope is caused by leakage current, the sudden voltage steps indicate photon events. Here, the detector is reset approximately every $ \SI{1}{\milli\second} $. \textbf{(b)} This zoomed-in plot shows three photon events in more detail.}
\end{figure}

The SDD sensor designed for SFA consists of 19 hexagonal-shaped pixels with an area of 26.6 mm² each, leading to a total sensitive area of 5.05 cm². In Figure \ref{fig:module_mechanics}(b) and (c) the sensor is depicted from both the entrance window and the front side, where the three HTRS ASICs are indicated.
Figure \ref{fig:sensor_scheme} shows schematically a cross-section of one pixel of the sensor from its center (drain contact) to the pixel edge (outer substrate contact os).
Every pixel has an integrated N-channel JFET in the center. Integrating this first amplification stage directly into each pixel has the advantage of reducing the stray capacitance and reducing microphony \cite{Lechner1996} \cite{Strueder1998}.
  \newline
A p-doped guard ring structure is implemented to decouple the FET from the bulk. This guard ring is biased from the inner guard ring contact (igr).
The 450 um thick silicon wafer is depleted from both sides, the planar p-doped back contact located on the photon entrance side, as well as the small p-doped drift rings on the front side are biased with a negative voltage with respect to the n-doped anode contact.
A drift field is generated in the sensitive volume by 
biasing each drift ring with a more negative voltage the further the ring is located towards the edge of a pixel.
This is accomplished with voltage-dividing resistors between every neighbouring drift ring. A bias voltage is only applied to the first ring (r1) and the last ring (rX).
In case of a photon impact in a pixel, electron-hole pairs are generated in the sensitive volume, where the number of generated charges is proportional to the energy of the photon.
The holes go towards the negatively biased back contact, whereas the electrons drift to the anode near the center of the pixel.  \newline
For most photon energies within the energy range of SFA, the photons will be absorbed near the entrance window and only higher-energetic X-rays have a significantly high probability of penetrating the sensor.
No matter the generation location, the drift time in vertical direction is negligibly small compared to the one in horizontal direction due to the sensor dimensions and the electric fields in it. 
A thin aluminium layer on top of the back contact is provided to block unwanted light in the visible energy range from entering the sensor. This so-called optical blocking filter is about $ \SI{100}{\nano\meter} $ thick and reduces the quantum efficiency, especially for X-rays below $ \SI{1}{\kilo\electronvolt} $, since the probability of those photons being absorbed in the aluminium layer is non-negligible \cite{Henke1993}. \newline
A feedback capacitance of $C_{f} = \SI{25}{\femto\farad}$ is integrated on the sensor in each pixel, connected to the anode and the output of a preamplifier present in each channel of the HTRS ASIC. This readout principle is referred to as a charge-sensitive amplifier (CSA), which is common for SDDs. Figure \ref{fig:readout} shows the detector's readout system, including the CSA.
If a charge is collected at the anode, it is integrated over the feedback capacitance, leading to a voltage increase in the CSA output signal OUT-PRE. Charge collection can occur due to leakage current in a pixel, the output response is then a linear voltage increase over time. In the case of an X-ray photon event, a significant number of electrons is generated in the active volume, causing a voltage step in the CSA output.
A monitoring signal allows to observe the OUT-PRE signal of the 8th channel of each ASIC. As an example, Figure \ref{fig:outsig}(a) shows the OUT-PRE signal of channel 8 of the HTRS ASIC S. \newline
To predict the OUT-PRE signal change in case of a photon event, we can first calculate the number of electron-hole pairs generated for a photon with the energy $ E_{ph} = \SI{5.895}{\kilo\electronvolt} $
\begin{equation}
	\label{eq:photonsgen}
	N_{el} = \frac{E_{ph}}{w} =  1602
\end{equation}
with the mean energy to create an electron-hole pair $ w=\SI{3.68}{\electronvolt} $ at $ \SI{0}{\degreeCelsius} $ \cite{Lowe2007}
and the energy  
$ E_{ph} = \SI{5.895}{\kilo\electronvolt} $ of the \textit{Mn-K}$_{\alpha} $ line.
This emission line is used here to verify the detector functionality.
The collection of those created electrons then causes an OUT-PRE signal change of
\begin{equation}
	\label{eq:outpsig}
	\Delta V \approx -\frac{Q}{C_{f}} = \frac{eN_{el}}{C_{f}} =\SI{10.3}{\milli\volt} \, ,
\end{equation}
where $ C_{f} $ is the feedback capacitance, and Q is the charge collected at the anode \cite{Lutz2007}.
This value is in good agreement with the observed voltage steps in Figure \ref{fig:outsig}(b). 
The further processing of the signal inside the HTRS ASIC is described in more detail in previous works. \cite{Bombelli2013,Altmann2024} In short, the preamp output signal passes a semi-gaussian shaper and a peak-stretching logic before being multiplexed to the HTRS ASIC output, as shown in Figure \ref{fig:readout}.
Time domain multiplexing is necessary to reduce the number of HTRS ASICs and also analog-digital converters (ADC) following these HTRS ASICs needed for the readout of all 19 pixels.
Since every HTRS ASIC features 8 channels, three ASICs are sufficient.
Due to the multiplexing, the time resolution deteriorates depending on the exact setting of the HTRS ASIC. The specification of which shaping time to choose will determine the setting of other parameters in the HTRS ASIC, e.g., the pile-up rejection phase duration \cite{Altmann2024}. A pile-up rejection logic is implemented in the HTRS ASIC to avoid determining wrong photon energies. This would happen if two photon events are too close in time to each other, and the resulting readout signals would affect each other's amplitude. The pile-up rejection logic discards such signals.
 \newline
The collection of charges at each pixel's anode is caused by bulk leakage current and photon-induced electrons. When the sensor is cooled down to temperatures like $ \SI{-30}{\degreeCelsius} $, this leakage current is much lower than at $ \SI{0}{\degreeCelsius} $. Thus, charge collection and therefore the OUT-PRE signal increase is usually dominated by photon events, even at moderate count rates. \newline 
In order to regularly drain the collected electrons from the anodes and therefore resetting the detector, a reset diode is implemented in every pixel. This diode is connected to the pixel's anode and is by default reverse biased with a negative voltage $ V_{lo} $. 
If we apply a short positive voltage pulse with voltage $ V_{hi} $
to the diode, all the electrons at the anode are drained. The values currently used are $ V_{lo}=\SI{-15}{\volt} $ and $ V_{hi}=\SI{5}{\volt} $, while the pulse duration is $ \SI{260}{\nano\second} $. Such a voltage pulse can be triggered by the HTRS ASICs via two different options. A periodic reset signal can be turned on with an adjustable period by adequately programming the HTRS ASIC. The other option is a comparator circuit that sends out a reset pulse every time the OUT-PRE signal passes a certain threshold voltage that can be adjusted with the HTRS ASIC programming.
The second option has the advantage of a self-adapting reset frequency depending on the photon intensity and leakage current.

\section{Detector Performance}
\label{sec:perf} 
\begin{figure} [ht]
	\begin{center}
		\begin{tabular}{c} 
			\includegraphics[width=0.97\linewidth]{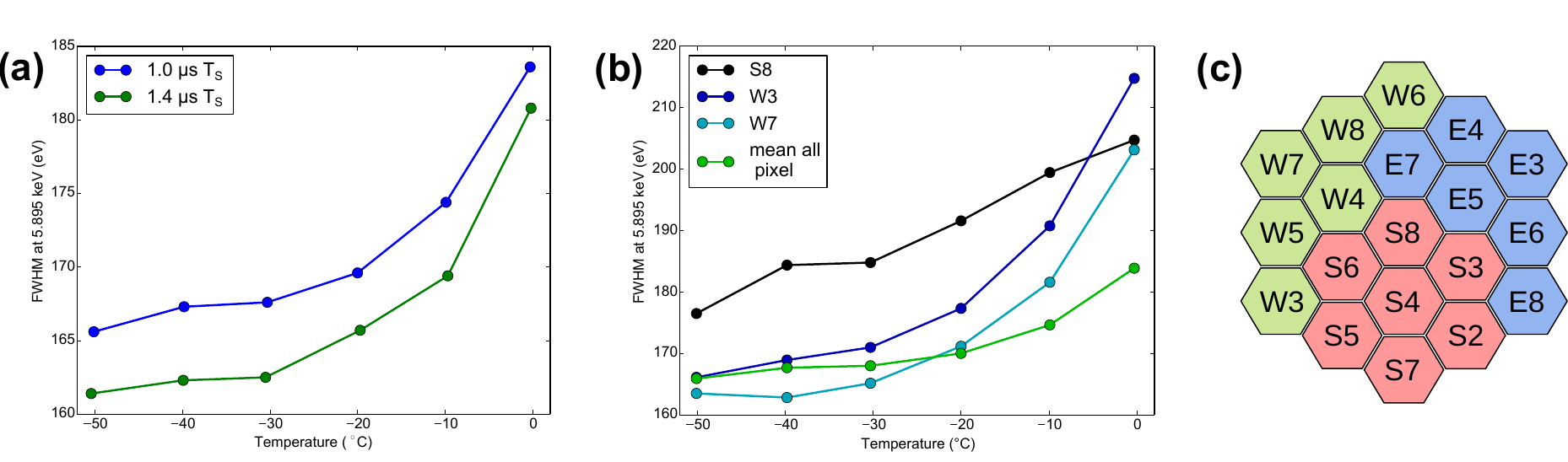}
		\end{tabular}
	\end{center}
	\caption[example] 

	{\label{fig:enres}\textbf{(a)} Mean energy resolution at $ \SI{5.895}{\kilo\electronvolt} $ for the two shaping times at different detector temperatures. \textbf{(b)} Energy resolution of three "bad" pixels at the shaping time $T_{S} = \SI{1.0}{\micro\second}$ compared to the mean of all pixels. \textbf{(c)} Sensor map, as seen from the front side, showing every pixel name. The name indicates to which HTRS ASIC (\textbf{W}est, \textbf{S}outh, \textbf{E}ast) and to which channel the pixel is connected.} 
\end{figure}

Applying the correct bias voltages to the SDD is crucial for the detector to operate properly. Table \ref{tab:bias_volt} shows a set of bias voltages used for the measurements presented later.
The given voltages of $ V_{bc}=\SI{-110}{\volt} $ for the back contact and $ V_{rX}=\SI{-186}{\volt} $ for the outer drift ring are sufficient to fully deplete the whole sensor.
The voltage difference between the innermost and outermost drift ring r1 and rX determines the strength of the drift field that electrons go through. A high drift field shortens the drift time of electrons, reducing the possibility of ballistic deficits.
If the inner ring voltage is too positive, the potential minimum is shifted towards the sensor front side (top side in Figure \ref{fig:sensor_scheme}). Then, charge collection can be incomplete since some electrons would get trapped between the first and second rings. In that case, the observed energy for X-rays would be lower than expected. This effect starts to become significant at a voltage of $ V_{r1}=\SI{-10}{\volt} $ and is avoided by selecting a more negative bias voltage.
\begin{table}[ht]
	\caption{\label{tab:bias_volt}Typical bias voltages used for measurements with the detector.} 
	\label{tab:Paper Margins}
	\begin{center}       
		\begin{tabular}{|c||>{\centering\arraybackslash}p{1.8cm}|>{\centering\arraybackslash}p{2cm}|>{\centering\arraybackslash}p{1.8cm}|>{\centering\arraybackslash}p{1.8cm}|>{\centering\arraybackslash}p{1.8cm}|} 	
			\hline
			\rule[-1ex]{0pt}{3.5ex} Voltage name & drain & back contact & back frame & ring 1 & ring X  \\
			\hline
			\rule[-1ex]{0pt}{3.5ex}  Voltage value & 4 V & -110 V & -115 V & -23 V & -186 V  \\
			\hline
		\end{tabular}
	\end{center}
\end{table} \newline
The HTRS ASIC has different shaping times available when processing signals. Two of those shaping times can fulfill the dead time requirement for eXTP, namely $ \SI{1.0}{\micro\second} $ and $ \SI{1.4}{\micro\second} $.
In Table \ref{tab:dead_time} we list the dead time of the detector as well as the relative dead time expected when observing a 1-Crab source.
The dead time given here includes the pile-up rejection phase duration programmed in the HTRS ASIC and also the shaper peaking time. This is because a potential pile-up rejection of a photon signal and its subsequent following signal have to be considered.
A more detailed description of the dead time in our system is given in a previous publication \cite{Altmann2024}.
The detector throughput or relative dead time can then be calculated for a given count rate by using the extendable dead time model. \cite{Knoll1999,Pomme2015} Observing a 1-Crab source with SFA would approximately lead to a photon count rate of $\SI{6.4}{\kilo cps}$ in the center pixel.
The obtained values for the relative dead time are satisfying the requirement of less than 5 \% at 1 Crab for the eXTP SFA.
\begin{table}[ht]
	\caption{\label{tab:dead_time}Two different shaping time settings in the HTRS ASIC and the resulting dead times as well as the relative dead time when observing a 1-Crab source.} 
	\begin{center}       
		\begin{tabular}{|c|c|>{\centering\arraybackslash}p{3.3cm}|} 	
			\hline
			\rule[-1ex]{0pt}{3.5ex}  shaping time & dead time & relative dead time at \newline 1 Crab ($\approx \SI{6.4}{\kilo cps}$)  \\
			\hhline{|=|=|=|}
			\rule[-1ex]{0pt}{3.5ex}  $\SI{1.0}{\micro\second}$ & $\SI{3.4}{\micro\second}$ & $ \SI{2.2}{\percent} $  \\
			\hline
			\rule[-1ex]{0pt}{3.5ex}  $\SI{1.4}{\micro\second}$ & $\SI{4.6}{\micro\second}$ & $ \SI{2.9}{\percent} $  \\
			\hline
		\end{tabular}
	\end{center}
\end{table} \newline
To further assess which shaping time satisfies the requirements for the eXTP mission, we determined the detector's energy resolution depending on the shaping time for different temperatures. 
Therefore, our prototype module was placed inside a vacuum chamber at a pressure below $ \SI{1d-6}{\milli\bar} $ where a radioactive Fe-55 source is located, emitting X-ray photons mainly at $ \SI{5.895}{\kilo\electronvolt} $ (\textit{Mn-K}$_{\alpha} $ line).
We evaluated the measurements with an analysis software using the ROOT based Offline Analysis (ROAn) framework \cite{Lauf2013}.
With the help of this software, the energy resolution at $ \SI{5.895}{\kilo\electronvolt} $, expressed as the full-width half maximum (FWHM), was determined for each measurement. Its mean of all pixels is plotted in Figure \ref{fig:enres}(a).
The $ \SI{1.4}{\micro\second} $ setting provides a better energy resolution.
At higher temperatures, this difference gets smaller, likely due to the leakage current increasing significantly and shot noise increasing linearly with both the leakage current and the shaping time.
If we consider the mean energy resolution at $ \SI{5.895}{\kilo\electronvolt} $, the requirement of a FWHM better than $ \SI{180}{\electronvolt} $ at $ \SI{6}{\kilo\electronvolt} $ for the eXTP mission is satisfied. Yet, some pixels are performing noticeably worse. Those are plotted in Figure \ref{fig:enres}(b), as well as the mean of all 19 pixels for comparison.
The pixel positions are shown in Figure \ref{fig:enres}(c), when looking at the front side comparable to Figure \ref{fig:module_mechanics}(c).
The pixel names are related to the corresponding HTRS ASIC channel they are connected to.  
Pixel W3 and W7 show a steep increase of the FWHM with rising temperature. This could indicate that an increased leakage current is present in those pixels.
The shot noise following from this leakage current is negligible at lower temperatures but becomes the dominant noise contribution compared to thermal and pink noise at high temperatures. \newline
A method to determine the leakage current is measuring the CSA output signal. In the absence of photons, the voltage increase is solely due to leakage current.
The HTRS ASIC only allows us to observe the OUT-PRE signal in the 8th channel of each HTRS ASIC. 
For the three pixels that can thus be monitored, 
Table \ref{tab:leakage_curr} shows the determined leakage current values at room temperature ($\SI{25}{\degreeCelsius}$). 
\begin{table}[ht]
	\caption{\label{tab:leakage_curr}Leakage current at room temperature determined from the CSA output ramp for different pixels.} 
	\begin{center}       					
		\begin{tabular}{|c||c|c|c|} 
			\hline
			\rule[-1ex]{0pt}{3.5ex} pixel name & W8 & S8 & E8 \\
			\hline
			\rule[-1ex]{0pt}{3.5ex} leakage current &  $\SI{0.43}{\nano\ampere\per\centi\meter\squared}$ & $\SI{0.31}{\nano\ampere\per\centi\meter\squared}$ & $\SI{0.44}{\nano\ampere\per\centi\meter\squared}$  \\
			\hline
		\end{tabular}
	\end{center}
\end{table}
The values are relatively high compared to similar devices also fabricated at the MPG Semiconductor Laboratory, which show a typical leakage current of around $ \SI{0.11}{\nano\ampere\per\centi\meter\squared} $ \cite{Gugiatti2022}.
Investigations are ongoing to explain this issue.
However, as long as the sensor is kept at low enough temperatures, like $ \SI{-30}{\degreeCelsius} $, the negative consequences of this leakage current are not too severe.
It has to be mentioned that this is the first detector module we built for eXTP. A more solid evaluation of the quality of the SDD sensors and the detector modules overall can only be made after manufacturing and testing of future modules. 
A way to qualitatively evaluate the leakage current in pixels not connected to an 8th HTRS ASIC channel is given by observing the reset periodicity. The pixel with the highest leakage current is the one that activates the threshold reset first because its OUT-PRE signal increases the fastest, as described earlier.
The HTRS ASIC allows us to deactivate specific channels, disabling them from triggering a reset pulse.
When channel W3 is deactivated, the reset period significantly increases. This indicates that W3 has the highest leakage current of all pixels. \newline
Another problematic pixel is S8, where between $ \SI{-50}{\degreeCelsius} $ and $ \SI{-10}{\degreeCelsius} $ the energy resolution is worse than in any other pixel.
This cannot be explained by a high leakage current since the determined value for S8 is lower than that of E8 and W8.
This is concerning because S8 is the central pixel of the sensor and therefore the most important one. The planned strategy for eXTP is to use only this pixel to collect all photons from a target source. All the surrounding pixels are then used to determine the background radiation.  
We have to closely monitor the performance, especially of pixel S8, in future modules to make sure it is in accordance with the eXTP requirements.

\section{summary}
A prototype detector module for the eXTP SFA was presented, and both the mechanical design and the sensor design were discussed.
The detector fulfills the eXTP requirements in terms of dead time and energy resolution for the shaping times $ \SI{1.0}{\micro\second} $ and $ \SI{1.4}{\micro\second} $ at temperatures up to $ \SI{-10}{\degreeCelsius} $, when the mean energy resolution of every pixel is considered. 
Still, some pixels show a poor performance.
More SDD modules are now manufactured, and we expect to get more reliable statistical information about leakage current and energy resolution by investigating those modules.

\acknowledgments     

Development and production of the SDD sensors for eXTP SFA are performed in collaboration between MPE and the MPG Semiconductor Laboratory (HLL). The authors want to thank Patricia Langer and Franz Soller for preparing the SDD detector module presented here.
This work was funded by the Max Planck Society.

\bibliography{bibliography_spie_paper}
\bibliographystyle{spiebib}

\end{document}